\documentclass[
    aps,
    prb,
    graphicx,
    superscriptaddress,
    twocolumb,
    reprint,
    floatfix,
    amsmath,
    amssymb
]{revtex4-2}

\usepackage{graphicx}
\graphicspath{{./figures}}
\usepackage[caption=false]{subfig}
\usepackage{dcolumn}
\usepackage{multirow}
\usepackage{physics}
\usepackage{bm}
\usepackage[ignoreunlbld,norefs,nocites]{refcheck}
\usepackage[ascii]{inputenc}
\usepackage{todonotes}
\newcolumntype{d}[1]{D{.}{.}{#1}} 
\newcolumntype{e}[1]{D{:}{}{#1}} 

\begin{document}

\title{Quantum Monte Carlo study of low-dimensional Fermi fluids of dipolar atoms}
\author{Clio Johnson}
\email{Contact author: d.k.johnson@lancaster.ac.uk, she/her/hers}
\author{Neil D.\ Drummond}
\affiliation{Department of Physics, Lancaster University, Lancaster, LA1 4YB, United Kingdom}
\author{James P.\ Hague}
\author{Calum MacCormick}
\affiliation{School of Physical Sciences, The Open University, Walton Hall, Milton Keynes, MK7 6AA, United Kingdom}
\date{\today}

\begin{abstract}
    Fermionic cold atoms in optical traps provide viable quantum simulators of correlation effects in electronic systems.
    For dressed Rydberg atoms in two-dimensional traps with out-of-plane dipole moments, a realistic model of the pairwise interaction is of repulsive dipolar $1/r^3$ form at long range, softened to a constant at short range.
    This study provides parametrizations of fixed-node diffusion Monte Carlo energy data for ferromagnetic (one-component) and paramagnetic (two-component) two-dimensional homogeneous Fermi fluids of interacting dipolar atoms.
    We find itinerant ferromagnetism to be unstable within our parameter spaces for dipolar interactions both with and without softening.
    Our parametrization of the energy as a function of density will enable density functional theory to support experimental studies of inhomogeneous fermionic cold atom systems.
\end{abstract}

\maketitle

\section{Introduction}

Cold atom systems offer a radically different approach to gain insight into quantum materials. In such systems, whole atoms moving in optical lattices or arrays of optical tweezers~\cite{Browaeys2020,Kaufman2021} form an analogous quantum system to electrons moving in the potential formed from a periodic array of nuclei. Cold atom systems can be highly controllable, with the caveat that the larger lengthscales associated with the lattice imply that energy scales are much smaller and the system needs to be cooled to very low temperatures on the order of nK to access quantum effects. Within these quantum analogues, a range of interaction strengths can be accessed allowing for the probing of weak to strong correlation regimes. 

There has been a lot of work on strong correlation systems such as Hubbard models and quantum spin systems demonstrating Mott insulating phases and antiferromagnetism~\cite{Bloch2008}, whereas quantum simulators for the weak correlation limit of quantum materials have been largely neglected. Weakly correlated quantum materials also have interesting phases and properties. These include Dirac materials such as graphene, the more general van der Waals heterostructures, and Weyl semimetals; systems in which edge states are important such as topological insulators and quantum spin hall insulators; band insulators and related charge density wave systems; itinerant ferromagnets and ferro- and piezo-electrics. The development of quantum simulators for such systems would be desirable.

Moreover, the intermediate correlation regime of quantum many-body systems is key to a variety of important yet poorly understood materials, such as the cuprate superconductors. In this regime, neither weak nor strong coupling perturbative approaches are reliable, the narrow band approximation associated with Hubbard's model breaks down~\cite{Hubbard1963}. The failure of narrow band approximations is particularly challenging as increasing numbers of bands and the interactions between them quickly overwhelm available techniques for strongly correlated systems, such as exact diagonalization, density matrix renormalization group (matrix product state) approaches, and various flavors of quantum Monte Carlo (QMC) calculation, such as determinant and path integral QMC\@. Issues with density functional theory (DFT) calculations relate to challenges around approximations to the exchange-correlation (XC) potential, which become progressively worse for intermediate to strong correlation.

Significant success has been achieved in the strong correlation limit with quantum simulators for Hubbard models, constructed using cold atoms in optical lattices interacting via \emph{short-ranged} Feshbach resonances, allowing both the strong and  weak coupling limits (unphysical when related to most materials, since bands become wide and overlap in such a limit, but achievable in a quantum simulator) of the model to be probed to examine the Mott transition~\cite{Bloch2008}. Recently, lattices of optical tweezers have been proposed to replace the optical lattice to attain greater control over the systems. A limitation of Feshbach resonances is that they emulate local $\delta$-function like interactions, consistent with Hubbard models, but at variance with weak and intermediate correlation systems where electron-electron interactions cannot be approximated as local.

Long-range interactions are key to exploring the weak and intermediate coupling regimes of quantum materials using quantum simulators. A key feature of the current work is the study of \emph{softened} dipole interactions relevant to dressed Rydberg atoms. Long-range interactions may be obtained using dressed Rydberg atoms~\cite{Gallagher2006,Johnson2010}, which have high principal quantum number excited states, resulting in a large dipole moment. Typically, a dressing scheme is used by exciting Rydberg states using light detuned from the transition. In this way, a superposition of ground and Rydberg states can be formed, and a higher level of tunability and stability is obtained. By tuning the dressing scheme, the dressed Rydberg atoms can be primed to interact via both dipole-dipole interactions and van der Waals interactions. Owing to higher-order corrections caused by the dressing scheme when atoms become close, the dipole-dipole interactions have the form
\begin{equation}%
\label{eq:dipole_dipole_interaction}
    v_{\rm dd}(r) = \frac{d^2}{r^3+r_0^3},
\end{equation}
where the dipole strength $d$ and softening parameter $r_0$ are constants, as shown in Ref.~\onlinecite{Hague2017}. For small interatomic separations, the Rydberg levels are shifted by the interaction. This shift decreases the excitation of the Rydberg levels by electromagnetic radiation leading to a saturated interaction (a mechanism that is analogous to the Rydberg blockade). 
We also note the use of systems of cold ions~\cite{Blatt2012} and quantum dots~\cite{Barthelemy2013} for quantum simulation. Interactions within quantum dot systems are typically local. The Coulomb interaction between cold ions is excellent for the study of spin systems, but is too large relative to the quantum simulator energy scales to probe anything away from the strong correlation limit~\cite{Hague2017}. The weak to intermediate correlation regime of quantum simulators is largely unexplored. 

This paper presents a continuum QMC study of a homogeneous fluid of dipolar atoms in two dimensions (2D), relevant to dressed Rydberg atoms in pancake traps~\cite{Henderson2009}.  Ground state energies, parameterized over a large range of interaction strengths are presented for the previously studied bare dipolar interaction~\cite{Comparin2019,Matveeva2012}, as well as softened dipolar interactions relevant to dressed Rydberg atoms. These parameterized energy data provide local density approximation (LDA) energy functionals that will allow DFT studies of inhomogeneous systems of dipolar atoms in optical traps.
We note that a DFT for cold-atoms interacting via Feshbach resonances was developed by Ma \textit{et al.}~\cite{Ma2012}. Fermionic quantum simulators with low dimensional geometries and dressed Rydberg atoms in 2D materials such as graphene and hexagonal boron nitride were studied by Hague and MacCormick~\cite{Hague2017}. The QMC calculations are also directly applicable to the changes of state that occur in cold atom traps containing dressed Rydberg atoms between dilute and dense limits; using fluid results we can explore the stability of itinerant ferromagnetism. The parameter space from Ref.~\onlinecite{Comparin2019} is extended in the bare dipolar case and the same analysis is completed with a softening parameter. Explorations of itinerant ferromagnetism have been performed for a number of different 2D homogeneous systems with short-range interactions as seen in Refs.~\onlinecite{Parish2013,Bombin2019,Pilati2021}.

This paper is organized as follows. In Sec.~\ref{sec:dft} we outline how our results can be used in DFT studies of inhomogeneous fermionic dipolar atomic fluids. In Sec.~\ref{sec:qmc} details of our QMC approaches are provided. In Sec.~\ref{sec:softened_dipoles} results for softened interactions [Eq.\ (\ref{eq:dipole_dipole_interaction})] appropriate for dressed Rydberg atoms of $^{43}$Ca are presented. In Sec.~\ref{sec:bare} results for the bare dipolar interaction ($r_0=0$) are presented and compared to existing data. We summarize and conclude in Sec.~\ref{sec:conclusions}.

A dataset of QMC calculations performed in service of this article is openly available~\cite{Johnson2025}.

\section{DFT for inhomogeneous dipolar atom systems}%
\label{sec:dft}

\subsection{Softened dipolar interaction}

Consider a finite inhomogeneous system of fermionic dipolar atoms of mass $m$ moving in 2D subject to an inhomogeneous external potential $v_{\rm ext}({\bf r})$, for example describing an optical trap.
Let $n({\bf r})$ be the atomic number density.
Suppose the interaction between the atoms is of the form of Eq.\ (\ref{eq:dipole_dipole_interaction}) with $r_0>0$.
By the Hohenberg-Kohn theorem~\cite{Hohenberg_1964}, the total energy can be written as
\begin{equation}
    E[n] = T_{\rm s}[n] + E_{\rm H}[n] + E_{\rm ext}[n] + E_{\rm xc}[n],
\end{equation}
where $T_{\rm s}[n]=-\frac{\hbar^2}{2m} \sum_i {\cal N}_i \int \phi_i^* \nabla^2 \phi_i \, {\rm d}^2{\bf r}$ is the kinetic energy of a noninteracting auxiliary system of atoms with number density $n({\bf r})=\sum_i {\cal N}_i |\phi_i({\bf r})|^2$, where $\{\phi_i\}$ is the set of orthonormal orbitals of the auxiliary system and $\{{\cal N}_i\}$ is the set of occupation numbers~\cite{Kohn_1965}.
The Hartree energy is $E_{\rm H}[n] = (1/2) \iint n({\bf r}) n({\bf r}') v_{\rm dd}(|{\bf r}-{\bf r}'|) \, {\rm d}^2{\bf r} \, {\rm d}^2{\bf r}'$ and the external potential energy is $E_{\rm ext}[n]=\int n({\bf r}) v_{\rm ext}({\bf r}) \, {\rm d}^2{\bf r}$.
Finally, $E_{\rm xc}[n]$ is the (unknown) exchange-correlation (XC) energy.
Note that atoms in the quantum simulator are treated as point-like particles in the following, a reasonable approximation on the $\sim 100\mathrm{nm}-1\mu\mathrm{m}$ lengthscales of optical lattices and arrays of optical tweezers.

Within the LDA~\cite{Kohn_1965},
\begin{equation}%
\label{eq:dft}
    E_{\rm xc}[n] \approx \int n({\bf r}){\cal E}_{\rm xc}(n({\bf r})) \, {\rm d}^2{\bf r},
\end{equation}
where ${\cal E}_{\rm xc}(n)$ is the XC energy per atom of a homogeneous atomic gas of number density $n$.
Unlike the usual approach for electrons, we parameterize ${\cal E}_{\rm xc}(n)$ directly instead of splitting it into exchange and correlation contributions, due to the lack of an analytical expression for the exchange energy of a homogeneous atomic gas. Nonetheless the exchange energy on its own can be found as is shown in Appendix~\ref{app:qmc:exchange_energy}.
Our parametrization of ${\cal E}_{\rm xc}(n)$ is presented in Sec.~\ref{sec:softened_dipoles}.
The extension to a local spin-density approximation for spin-polarized systems is straightforward using a suitable interpolation~\cite{Barth_1972} between our paramagnetic and ferromagnetic XC functionals.

The Hartree energy is inaccurate in low-density regions (see Appendix~\ref{app:qmc:hartree_energy}), so that we rely heavily on the (approximate) XC energy to correct it.
However, low-density regions make a relatively small contribution to the total energy.

\subsection{Bare dipolar interaction}

Constructing a DFT methodology for a bare dipolar interaction ($r_0=0$) is more challenging, because the integral for the Hartree energy diverges.
A possible DFT approach for studying inhomogeneous atomic systems with bare dipolar interactions is to use the softened dipolar interaction LDA functional of Eq.\ (\ref{eq:dft}) together with an additional LDA term
\begin{equation}
    E_{\rm bc}[n] \approx \int n({\bf r}) \left[{\cal E}_{\rm bare}(n({\bf r})) - {\cal E}_{\rm soft}(n({\bf r}))\right] \, {\rm d}^2{\bf r}
\end{equation}
to correct for the difference in the interaction, where ${\cal E}_{\rm bare}(n)$ and ${\cal E}_{\rm soft}(n)$ are the energies per atom of bare dipolar and softened dipolar homogeneous atomic gases of number density $n$.
This is a potentially significant uncontrolled approximation, especially in regions of high atomic density.

\section{QMC for dipolar fermions}%
\label{sec:qmc}

\subsection{Units}

\subsubsection{Softened dipolar interaction}

Consider a homogeneous 2D fluid of fermionic atoms of mass $m_{\rm d}$ interacting via the potential given in Eq.\ (\ref{eq:dipole_dipole_interaction}).
Two separate systems of natural units are employed throughout this paper, given explicitly in terms of their composite constants.
The first (a.u.*)\ is used for softened interaction calculations as an aid to parameterizing ground state energies in terms of $r_{\rm s}$, the mean radius containing one atom, defined in relation to number density $n$ via $\pi r_{\rm s}^2 n = 1$ a.u.*
(Throughout we treat the density parameter $r_{\rm s}$ as a quantity with dimensions of length.)
The system of units is defined by $d^2=\hbar=m_{\rm d}=1$ a.u.* Hence the unit of length is ${m_{\rm d}}d^2/\hbar^2$, the unit of mass is $m_{\rm d}$, and the unit of energy is $\hbar^6/(m_{\rm d}^3d^4)$.

We parameterize the energy as a function of $r_{\rm s}$.
In this system of units, the kinetic energy and long-range interaction terms in the Hamiltonian go as $1/r_{\rm s}^2$ and $1/r_{\rm s}^3$, respectively.  Hence, unlike the case for a homogeneous electron gas, $r_{\rm s}\to\infty$ is the weakly interacting limit and $r_{\rm s}\to 0$ is the strongly interacting limit.
The softening parameter $r_0 = 0.6397686$ a.u.*\ used in our work is appropriate for $^{43}$Ca with principal quantum number $32$~\cite{Hague2017}.  In this system of atomic units the numerical value of $r_0$ is independent of $r_{\rm s}$.

\subsubsection{Bare dipolar interaction}

For the special case of the $r_0 = 0$ bare dipolar interaction, to allow easy comparison with previous work, we have fixed chosen $r_{\rm s}$ to be the unit of length.  The set of units are defined by $r_{\rm s}=\hbar=m_{\rm d}=1$ a.u., and units in this system are abbreviated to a.u.
In this case, the value of $d^2$ characterizes interaction strength. The unit of dipole interaction strength is $\hbar^2r_{\rm s}/m_{\rm d}$, the unit of length is $r_{\rm s}$, the unit of mass is $m_{\rm d}$, and the unit of energy is $\hbar/(m_{\rm d}r_{\rm s}^2)$.

\subsection{Methodology}

Pairwise interactions of the form shown in Eq.~(\ref{eq:dipole_dipole_interaction}) were implemented in the \textsc{casino} QMC code~\cite{Needs2020} for performing variational Monte Carlo (VMC) and diffusion Monte Carlo (DMC) calculations.
In the VMC method we take the expectation value of the Hamiltonian operator with respect to a trial wave function $\psi_{\rm T}$. The trial wave function contains free parameters that are determined by minimizing the energy expectation value or the variance of the energy.  In the DMC method we simulate drift, diffusion, and branching/dying processes governed by the Schr\"{o}dinger equation in imaginary time in order to project out the ground state component of the trial wave function. We maintain fermionic antisymmetry by fixing the complex phase of the DMC wave function.

Equation~(\ref{eq:dipole_dipole_interaction}) is applicable to atoms moving in one or two dimensions, with dipole moments perpendicular to the line or plane to which the atoms are confined.
The systems studied here are one-component (ferromagnetic) and two-component (paramagnetic) fermionic dipolar monolayers.
Two-dimensional ferromagnetic fluid and crystal phases of fermionic atoms interacting via the bare dipolar potential have been studied in the past by Matveeva and Georgini~\cite{Matveeva2012}.
Paramagnetic fluids have been studied more recently by Comparin \textit{et al.}~\cite{Comparin2019}, whose work compares para- and ferromagnetic phases to assess for itinerant ferromagnetism.

The Hamiltonian for a system of $N$ dipolar atoms moving in a 2D periodic simulation cell is
\begin{equation}
    \hat{H} = -\frac{1}{2}\sum\limits_{i=1}^{N}\nabla^2_i + \sum\limits_{i=1}^{N-1}\sum\limits_{j=i+1}^{N}\sum\limits_{\vb{R}_{\rm s}} v_{\rm dd}(|{\bf r}_i-{\bf r}_j+\vb{R}_{\rm s}|) + \frac{Nv_{\rm M}}{2},
\end{equation}
where $\vb{r}_i$ are atom positions, $\{\vb{R}_{\rm s}\}$ are simulation-cell lattice points, and the Madelung constant $v_{\rm M}$ is the interaction energy between each atom and its own periodic images.
In practice the sum over interaction energies is only evaluated explicitly up to a radius $\vqty{\vb{R}_{\rm s}}<R_{\rm c}$, beyond which the absolutely convergent tail of the interaction term is approximated by an integral~\cite{Drummond2011},
\begin{equation}%
    \begin{split}
        & \sum_{{\bf R}_{\rm s}} v_{\rm dd}(|{\bf r}+{\bf R}_{\rm s}|) \\ \approx & \sum\limits_{\vb{R}_{\rm s} : R_{\rm s}<R_{\rm c}} v_{\rm dd}(|\tilde{\bf r}+{\bf R}_{\rm s}|) + \frac{1}{A} \int_{R_{\rm s}>R_{\rm c}} v_{\rm dd}(|\tilde{\bf r}+{\bf R}_{\rm s}|) \, {\rm d}^2{\bf R}_{\rm s} \\ \approx & \sum\limits_{\vb{R}_{\rm s} : R_{\rm s}<R_{\rm c}} v_{\rm dd}(|\tilde{\bf r}+{\bf R}_{\rm s}|) + \frac{2\pi d^2}{A R_{\rm c}} \left(1 + \frac{3\tilde{r}^2}{4R_{\rm c}^2}\right) + O(R_{\rm c}^{-3}).
    \end{split} \label{eq:dd_int}
\end{equation}
Here $A=\pi r_{\rm s}^2 N$ is the area of the simulation cell and $\tilde{\bf r}$ is the closest image distance between interacting dipoles.  Likewise, the Madelung constant is in practice evaluated as
\begin{equation} v_{\rm M} \approx \sum\limits_{\vb{R}_{\rm s} : 0<R_{\rm s}<R_{\rm c}} v_{\rm dd}(R_{\rm s}) + \frac{2\pi d^2}{A R_{\rm c}} + O(R_{\rm c}^{-3}). \label{eq:dd_vM} \end{equation}
We have used 119 stars of lattice vectors in the explicit sum in Eqs.\ (\ref{eq:dd_int}) and (\ref{eq:dd_vM}).  Hence $R_{\rm c}$ grows with system size $N$, so that the error due to the approximation of the infinite lattice sum can be regarded as another source of finite-size error.
The error introduced by using this integral correction instead of an infinitely large sum is significant compared to our QMC precision in high density calculations, but small when compared to relevant energy scales such as the XC energies calculated in Sec.~\ref{sec:softened_dipoles}.
Analysis of these errors can be found in Appendix~\ref{app:qmc:err_sum}

At long range the dipolar interaction of Eq.\ (\ref{eq:dipole_dipole_interaction}) reduces to $d^2/r^3$, and hence the expressions for the integral corrections to the lattice sums in Eqs.\ (\ref{eq:dd_int}) and (\ref{eq:dd_vM}) are evaluated using $d^2/r^3$.  The fractional error in the integral approximation due to the assumption that $r_0=0$ is less than $2 \times 10^{-5}$ in all our calculations.

DMC energies were obtained using two finite imaginary time step calculations, the majority of which were at a ratio of $1:4$.
The DMC energies were then extrapolated to the zero imaginary time step limit.

The use of a nondivergent pseudopotential has been explored by Whitehead and Conduit~\cite{Whitehead2016} to facilitate QMC studies of dipolar atomic gases.  However, by using appropriate trial wave function forms we have been able to calculate energies to high precision without the need to introduce a pseudopotential approximation.

All DMC fluid calculations were twist-averaged, usually over 1,000 twists.
Precise DMC energies were obtained through the use of the free-atom kinetic energy per atom as a correlator when twist averaging.

\subsection{Trial wave functions}

Two different trial wave functions were used for dipolar fluid calculations.
Taking $\vb{R}$ to be the $2N$-dimensional vector of atomic positions $(\vb{r}_1,\vb{r}_2,\ldots,\vb{r}_N)$, the simpler Slater-Jastrow (SJ) wave function is
\begin{equation}
    \psi_{\rm T}\pqty*{\vb{R}} = S\pqty*{\vb{R}}\exp\pqty*{J\bqty*{\vb{R}}}.
\end{equation}
Here, $S\pqty*{\vb{R}}$ is a Slater determinant of single-atom orbitals providing the requisite fermionic antisymmetry, with one determinant per spin component; $\exp(J)$ is an optimizable Jastrow factor~\cite{Drummond2004} accounting for correlation.
The Jastrow factor is a sum of two- and three-body polynomial terms $u$ and $H$, respectively, as well as a plane-wave expansion over two-body separations $p$ included to describe long-range correlations and hence to facilitate the extrapolation of energies per atom to the thermodynamic limit.

Slater-Jastrow-backflow (SJB) wave functions include a backflow displacement~\cite{LopezRios2006}.
The following form was used
\begin{equation}%
\label{eq:trial_wavefunction}
    \psi_{\rm T}\pqty*{\vb{R}} = S\bqty*{\vb{X}\pqty*{\vb{R}}}\exp\pqty*{J\bqty*{\vb{R}}},
\end{equation}
where $\vb{X}$ is a set of quasiparticle coordinates $\pqty{\vb{x}_1,\vb{x}_2,\ldots\vb{x}_N}$ displaced from the original atom coordinates as $\vb{x}_i = \vb{r}_i + {\bm\xi}_i\pqty*{\vb{R}}$.
This displacement is dependent on all the atom coordinates and is a sum of two-body polynomial and plane-wave expansions ${\bm \eta}$ and ${\bm \pi}$ respectively.
The main purpose of backflow is to allow a variational optimization of the nodal surface of the trial wave function.

In a DMC calculation, in the infinite imaginary time limit, the fixed-node ground state component of the trial wave function is projected out.
Practically this means that the fixed-node ground state energy for a DMC simulation is improved by a well-optimized backflow displacement, whereas a well-optimized Jastrow factor facilitates convergence without affecting the fixed-node ground state energy.
Backflow displacements have been used in our bare ($r_0 = 0$) dipolar calculations, because the interaction is relatively strong; e.g., backflow affects the residual divergent contributions to the local energy $(\hat{H}\psi_{\rm T})/\psi_{\rm T}$ at coalescence points discussed in Appendix~\ref{app:qmc:short_range}.
The effect of backflow on the DMC energy can be seen in Fig.~\ref{fig:d2_1.0_finite_size}.

\subsection{Parametrization of the energy and exchange-correlation energy}

The Hartree{--}Fock theory of the (softened) dipolar atomic gas is presented in Appendix~\ref{app:qmc:hartree-fock}. Analytical expressions are given for the noninteracting kinetic per particle and the Hartree energy per particle, and the behavior of the exchange energy per atom is discussed.

For use in DFT calculations we provide parametrizations of the total energy per atom as a function of interaction strength $d^2$ (for the bare dipolar interaction) or density parameter $r_{\rm s}$ (for the softened interaction) and, in the case of the softened interaction, we provide a parametrization of the XC energy per atom, defined as the difference between the total energy and the sum of the noninteracting kinetic energy and the Hartree energy, as a function of density parameter $r_{\rm s}$.

\subsection{Finite-size effects}

\begin{figure}[htbp!]
    \centering
    \subfloat[Paramagnetic system.]{%
        \includegraphics[width=0.95\columnwidth]{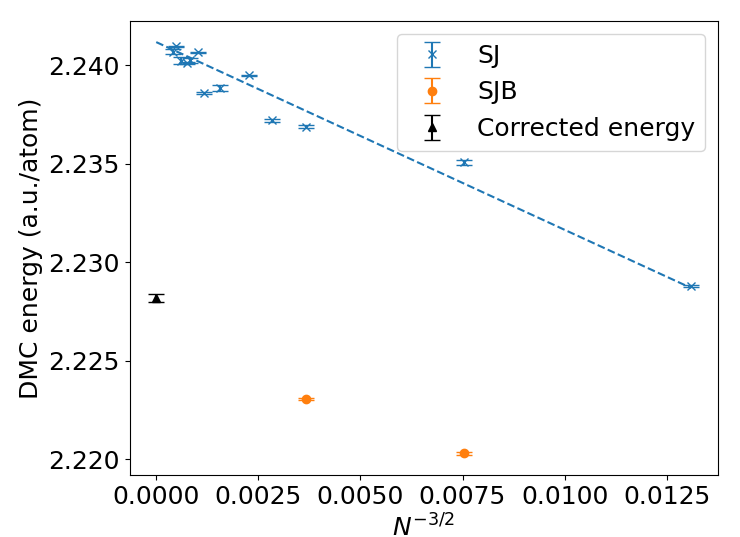}
    }\\
    \subfloat[Ferromagnetic system.]{%
        \includegraphics[width=0.95\columnwidth]{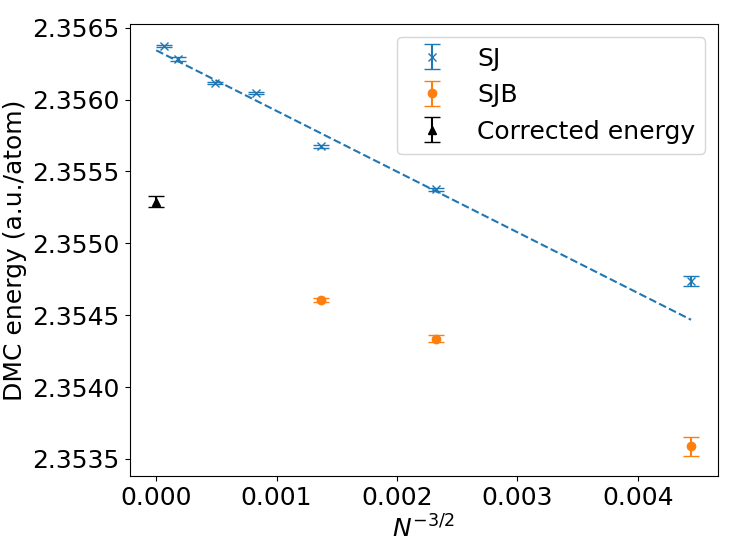}
    }
    \caption{DMC total energy per atom in bare ($r_0 = 0$) dipolar calculations performed at $d^2=1$ a.u.\ with both SJ and SJB wave functions against system size. The backflow-corrected DMC energy per atom in the thermodynamic limit $E = E_\infty^{\text{SJ}} + B_\infty$ is also shown.}%
    \label{fig:d2_1.0_finite_size}
\end{figure}

All fluid simulations were performed in finite square cells subject to twisted periodic boundary conditions.
As such, systematic finite-size errors dependent on the number of atoms in the simulation cell occur.
A straightforward linear extrapolation in energies per atom from several finite calculations was performed, using the scaling law
\begin{equation}%
\label{eq:energy_extrapolation}
    E_N = E_\infty + \frac{k}{N^{3/2}},
\end{equation}
where $E_N$ is the energy per atom in an $N$-atom cell, and $E_\infty$ and $k$ are parameters determined by fitting.
This scaling law is derived in Appendix~\ref{app:qmc:finite_size}.
In addition, there are smaller finite-size effects due to residual momentum-quantization errors after twist averaging and due to the truncation of lattice sums of interacting dipoles.
Equation~\eqref{eq:energy_extrapolation} has been shown to hold in both paramagnetic and ferromagnetic calculations in the bare dipolar case, as can be seen in Fig.~\ref{fig:d2_1.0_finite_size}.
The noise in the paramagnetic bare dipolar results in Fig.~\ref{fig:d2_1.0_finite_size}(a) is a consequence of the presence of distinguishable atom pairs, which have a sharper whole-wave-function dependent divergence in local energy as explained in Appendix~\ref{app:qmc:short_range}.
Since energies obtained here are from suitably converged DMC calculations, this noise cannot be attributed to poorly optimized Jastrow factors. On the other hand, momentum quantization effects could in principle affect the optimized backflow function and hence SJB-DMC energy.
Furthermore, the considerably lower degree of noise present in ferromagnetic results lends credence to the noise being a result of local energies diverging at different rates at coalescence points.
The systematic finite-size error in the ferromagnetic fluid energy per atom is an order of magnitude smaller than in the paramagnetic fluid.

In all cases atom numbers $N$ were chosen from the set of so-called ``magic'' numbers with closed shell occupancies in reciprocal space when the simulation cell is subject to non-twisted periodic boundary conditions.
In all cases our optimization calculations were performed at the Baldereschi mean-value point of the simulation cell~\cite{Baldereschi1973}.

In softened ($r_0 > 0$) dipolar calculations, results were extrapolated to the thermodynamic limit from system sizes of $N = 98$ and $122$ for paramagnetic systems and $N = 97$ and $121$ for ferromagnetic systems.
In bare ($r_0 = 0$) dipolar calculations, SJ calculations at system sizes $N = 42$ and $90$ were used for extrapolation in paramagnetic fluids. In ferromagnetic fluids system sizes of $N = 37$, $81$, and $113$ were used.
Various system sizes were examined for $d^2 = 1$ a.u.\ to probe the finite-size behavior in the bare dipolar case as seen in Fig.~\ref{fig:d2_1.0_finite_size}.

Differences in energy per atom due to the inclusion of backflow were calculated as $B_N = E_N^{\text{SJ}} - E_N^{\text{SJB}}$, and extrapolated to infinite system size using Eq.~\eqref{eq:energy_extrapolation}.
Separately extrapolating the backflow correction and the SJ-DMC energy to the thermodynamic limit allows much larger system sizes to be used for the latter, which is the dominant contribution to the total energy.
System sizes $N = 26$ and $42$ were used for the calculation of $B_\infty$ in paramagnetic fluids, whereas for ferromagnetic fluids, system sizes $N = 37$ and $57$ were used.
The extrapolation of results, especially in paramagnetic fluids, introduces unquantified quasirandom finite-size noise into the final energy per atom; however our quoted error bars purely represent the random errors due to QMC sampling.
The paramagnetic system sizes were chosen to have relatively far apart $1/N$ values to reduce the overall effect of noise on these calculations.
Whilst this noise is considerably lower in ferromagnetic calculations, backflow corrections are also less important.

\begin{figure}[htbp!]
    \centering
    \includegraphics[width=0.95\columnwidth]{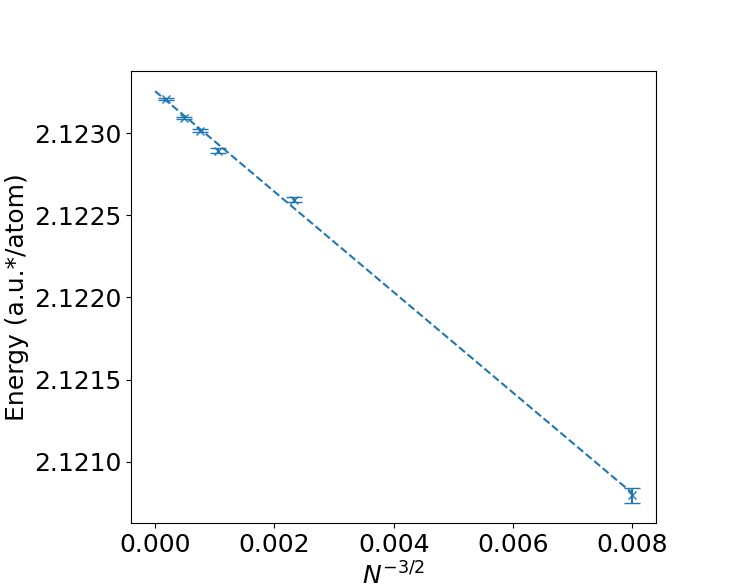}
    \caption{SJ-DMC total energies per atom of ferromagnetic softened dipolar systems calculated at $r_{\rm s}=1$ a.u.* plotted against system size. A finite-size fit is included. System sizes vary from $N=25$ to $N=317$.}%
    \label{fig:soft_rs_1.0_finite_size}
\end{figure}

In the softened backflow case, systemic finite-size effects were probed by gathering data at $r_{\rm s} = 1$ a.u.* in a ferromagnetic system and then extrapolating to the thermodynamic limit using the same leading order finite-size analysis derived in Appendix~\ref{app:qmc:finite_size}.
This can be seen in Fig.~\ref{fig:soft_rs_1.0_finite_size}.

\section{Softened dipolar interaction results}%
\label{sec:softened_dipoles}

The use of a softening parameter ($r_0>0$) results in faster DMC calculations due to the local energies becoming finite at coalescence points.
The energies per atom of paramagnetic and ferromagnetic fluids are summarized in Table~\ref{tab:softened_dipoles:energies}.

\begin{table*}[htbp!]
    \centering
    \caption{\label{tab:softened_dipoles:energies} Softened dipolar total energies per atom $E$ extrapolated to the thermodynamic limit against density parameter $r_{\rm s}$. These are presented alongside XC energies per atom $E_{\text{XC}}$, and Hartree-Fock energies per atom $E_{\text{HF}}$.}
    \begin{tabular}{d{2}d{9}d{9}d{9}d{9}d{9}d{9}}
        \hline\hline
        \multicolumn{1}{l}{\multirow{2}{*}{$r_{\rm s}$ (a.u.*)}} & \multicolumn{3}{c}{Paramagnetic} & \multicolumn{3}{c}{Ferromagnetic} \\
        & \multicolumn{1}{c}{$E$ (a.u.*/atom)} & \multicolumn{1}{c}{$E_{\text{XC}}$ (a.u.*/atom)} & \multicolumn{1}{c}{$E_{\text{HF}}$ (a.u.*/atom)} & \multicolumn{1}{c}{$E$ (a.u.*/atom)} & \multicolumn{1}{c}{$E_\text{XC}$ (a.u.*/atom)} & \multicolumn{1}{c}{$E_{\text{HF}}$ (a.u.*/atom)} \\
        \hline
        0.05 & 954.13306(6) & -1.89006(6)  & 954.19521 & 1154.12577(4) & -1.89734(4)   & 1154.17123 \\
        0.1  & 237.2383(6)  & -1.7675(6)   & 237.2607  & 287.2005(2)   & -1.8053(2)    & 287.2120   \\
        0.25 & 36.7046(3)   & -1.5363(3)   & 36.7603   & 44.6048(2)    & -1.6361(2)    & 44.6273    \\
        0.5  & 8.4514(1)    & -1.1088(1)   & 8.5442    & 10.23645(8)   & -1.32378(8)   & 10.27890   \\
        1.0  & 1.8056(1)    & -0.5844(1)   & 1.9114    & 2.12332(4)    & -0.76674(4)   & 2.16025    \\
        2.5  & 0.21284(5)   & -0.16956(5)  & 0.26736   & 0.25059(5)    & -0.21182(5)   & 0.25697    \\
        5.0  & 0.04295(7)   & -0.05265(7)  & 0.06262   & 0.052344(5)   & -0.063258(5)  & 0.053267   \\
        10.0 & 0.00901(2)   & -0.01488(2)  & 0.01507   & 0.011629(1)   & -0.017271(1)  & 0.011734   \\
        25.0 & 0.001153(6)  & -0.002671(6) & 0.002353  & 0.0017090(2)  & -0.0029151(2) & 0.0017139  \\ \hline\hline
    \end{tabular}
\end{table*}

\begin{figure}[htbp!]
    \centering
    \subfloat[Paramagnetic.]{%
        \includegraphics[width=0.95\columnwidth]{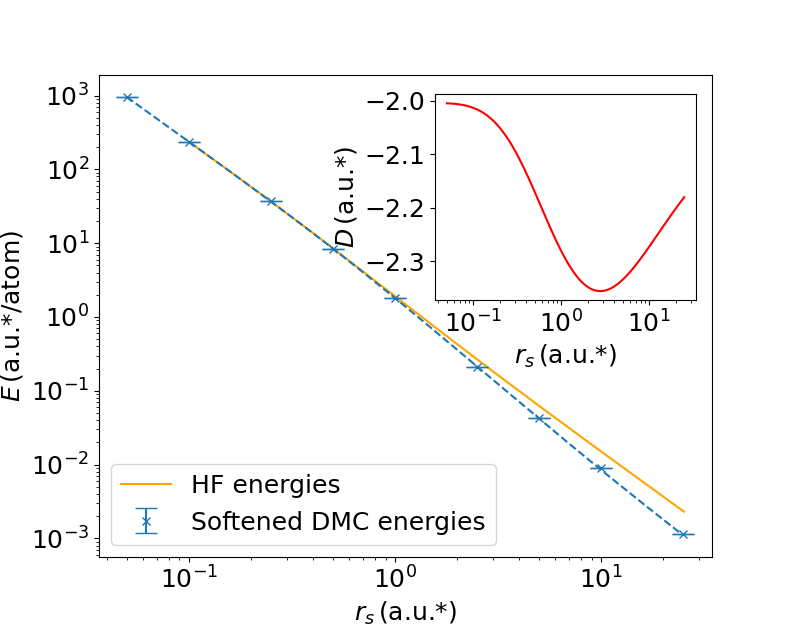}
    } \\
    \subfloat[Ferromagnetic.]{%
        \includegraphics[width=0.95\columnwidth]{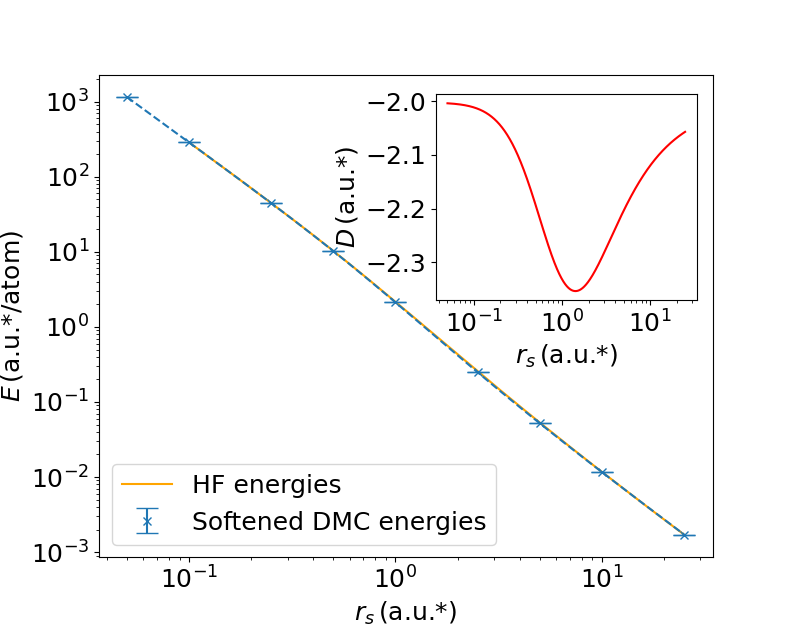}
    }
    \caption{DMC energy per atom and Hartree-Fock (HF) energies are plotted against the density parameter $r_{\rm s}$ in paramagnetic and ferromagnetic softened dipolar Fermi fluids. The logarithmic gradients $D={\rm d}\ln(E)/{\rm d}\ln(r_{\rm s})$, representing the exponent of decay of the energy per atom with $r_{\rm s}$, are presented in the insets.}%
    \label{fig:softened_dipoles:fit}
\end{figure}

Considering the $1/r^3$ form the softened interaction takes between dipoles at long range, in the low density (large $r_{\rm s}$) limit, the interaction energy will become negligible, resulting in a decay of $E \propto 1 / r_{\rm s}^2$.
At high density both the kinetic energy and the interaction energy (see Appendix~\ref{app:qmc:hartree_energy}) go as $1/r_{\rm s}^2$.
As seen in Fig.~\ref{fig:softened_dipoles:fit}, the energy per atom does indeed go as $E \propto 1 / r_{\rm s}^2$ at both high and low densities, as better visualised by the logarithmic gradient $D$ plotted in the inset, representing the exponent of the decay in $E \propto r_{\rm s}^D$.
To fit to the data found in Table~\ref{tab:softened_dipoles:energies}, the following Pad{\'e}-like function was used:
\begin{equation}%
\label{eq:softened_pade-like}
    E\pqty{r_{\rm s}} = \frac{a_0 + a_1\sqrt{r_{\rm s}} + a_2r_{\rm s} + a_3r_{\rm s}\sqrt{r_{\rm s}} + a_4r_{\rm s}^2}{r_{\rm s}^2+b_0r_{\rm s}^3 + pa_4r_{\rm s}^4},
\end{equation}
where $\{a_i\}$ and $b_0$ are fitting parameters.
$p$ is 2 for a paramagnetic fluid and 1 for a ferromagnetic fluid, and fixes the low-density asymptote to the kinetic energy given in Appendix~\ref{app:qmc:hartree-fock:kinetic}.

After fitting, in the paramagnetic case the $\chi^2$ value is $25.1$ per degree of freedom, and the total root-mean-square (RMS) error is $4.12\times 10^{-3}$ a.u.*/atom.
In the ferromagnetic case the $\chi^2$ value is $93.0$ per degree of freedom, with the total RMS error being $1.05\times 10^{-2}$ a.u.*/atom.
These $\chi^2$ values are high, but this is an expected deviation, since the error bars used only represent QMC error, which is very small.
The remaining error results from non-leading-order finite-size effects (including those seen in Appendix~\ref{app:qmc:err_sum}), which are not accounted for by the finite-size extrapolation used here.
Residuals for the first four results are all suitably small, but remain larger for higher $r_{\rm s}$ results, which is where most of the remaining residual comes from.
Optimized fitting parameters are reported in Table~\ref{tab:softened_energies:fit}

\begin{table}
    \centering
    \caption{Fitting parameters for the total DMC energy parametrization from Eq.~(\ref{eq:softened_pade-like}) optimized for para- and ferromagnetic softened dipolar systems.  All parameters are in a.u.*}
    \begin{tabular}{ld{8}d{8}}
        \hline\hline
        Param. & \multicolumn{1}{c}{Paramagnetic} & \multicolumn{1}{c}{Ferromagnetic} \\
        \hline
        $a_0$ & 2.43448037  & 2.91360411 \\
        $a_1$ & -0.36560564 & -0.20857602 \\
        $a_2$ & 11.47517594 & 2.77908742 \\
        $a_3$ & 3.02564825  & 0.2780977 \\
        $a_4$ & 2.45802893  & 1.72072964 \\
        $b_0$ & 4.62670384  & 0.80132663 \\ \hline\hline
    \end{tabular}%
    \label{tab:softened_energies:fit}
\end{table}

At larger $r_{\rm s}$ values, the calculations become increasingly expensive.
In order to accurately reflect the short-range behavior introduced by the softening parameter, the DMC time steps chosen were chosen such that the root-mean-square distance diffused by each atom in a single time step was considerably smaller than $r_0$.
On the other hand, the numbers of DMC equilibration and statistics accumulation iterations were large enough that walkers were able to diffuse across the entirety of the simulation cell.
Where this became computationally infeasible in the high $r_{\rm s}$ limit, the former condition of capturing the shortest lengthscale of the system was preferred.
We investigated whether it was possible to neglect the short lengthscale of $r_0$ in the high $r_{\rm s}$ limit by finding DMC time step biases in the less noisy ferromagnetic case.
In the $r_{\rm s} = 10$ a.u.* case, the difference between using $\Delta\tau_1 = 0.04$ a.u.* and $\Delta\tau_2 = 0.01$ a.u.* as opposed to $\Delta\tau_1 = 0.64$ a.u.* and $\Delta\tau_2 = 0.16$ a.u.* is $2.9(8)\times 10^{-6}$ a.u.*/atom, which is statistically significant.
As a result, calculations cannot be shortened for $r_{\rm s} = 10$ a.u.* calculations
In the $r_{\rm s} = 25$ a.u.* cases, time step data was gathered and the bias observed from using $\Delta\tau_1 = 0.4$ and $\Delta\tau_2 = 0.1$ as opposed to $\Delta\tau_1 = 1.6$ and $\Delta\tau_2 = 0.4$ is $2(1)\times 10^{-7}$, or statistically insignificant.

We find no evidence of a Bloch phase transition and subsequent itinerant ferromagnetism in softened dipolar results within the considered parameter space.
Hartree{--}Fock theory predicts a Bloch transition at $r_{\rm s} = 2.0334$ a.u.* with the ferromagnetic fluid favorable in the low-density limit.
This unphysical result is due to the erroneous divergence of Hartree-Fock theory in paramagnetic systems shown in Appendix~\ref{app:qmc:hartree_energy}.

\begin{figure}[htbp!]
    \centering
    \subfloat[Paramagnetic.]{%
        \includegraphics[width=0.95\columnwidth]{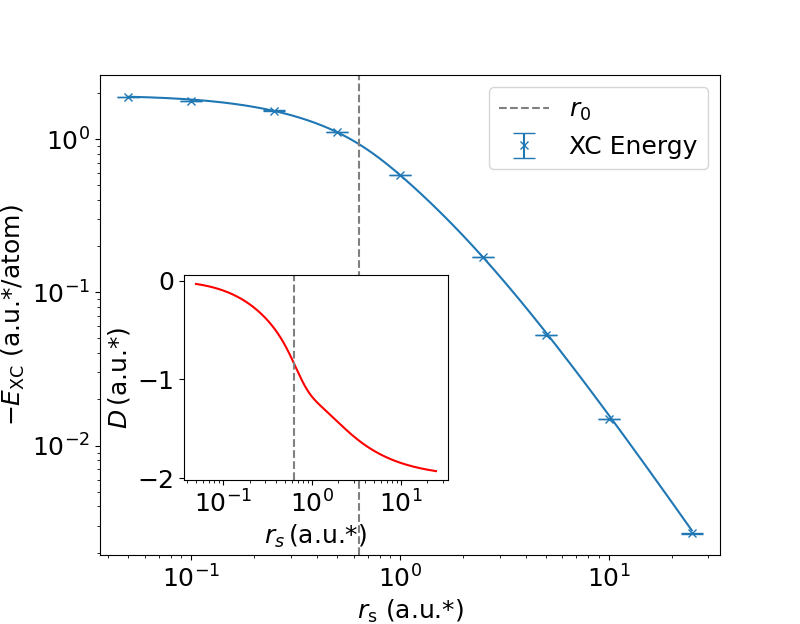}
    } \\
    \subfloat[Ferromagnetic.]{%
        \includegraphics[width=0.95\columnwidth]{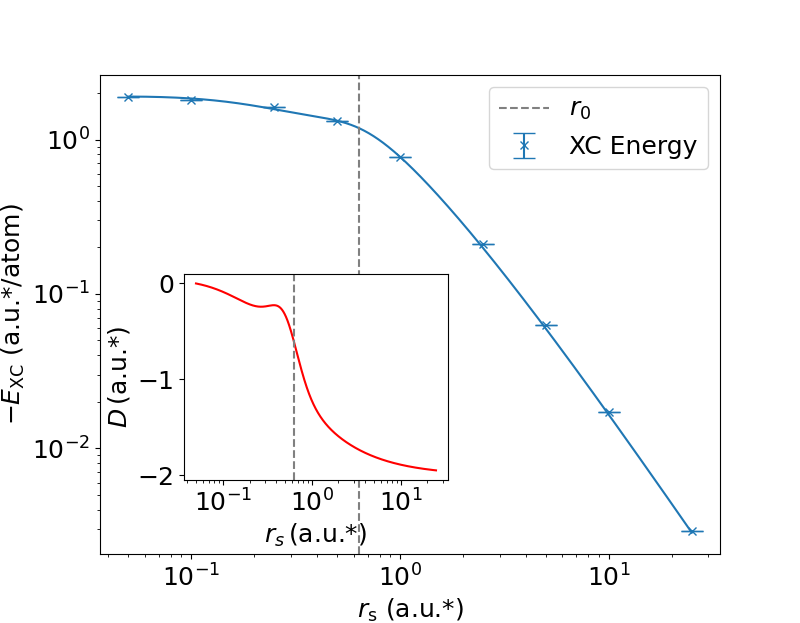}
    }
    \caption{XC energies per atom against density parameter $r_{\rm s}$ in para- and ferromagnetic softened dipolar Fermi fluids. The logarithmic gradient $D={\rm d}\ln(-E_{\rm XC})/{\rm d}\ln(r_{\rm s})$ of the fitted function is presented in the inset.}%
    \label{fig:softened_dipoles:xc}
\end{figure}

Fits of the XC energies per atom ($E_{\rm XC}$) have been produced with the Pad{\'e}-like fitting function
\begin{equation}%
\label{eq:xc_pade-like}
    E_{\rm XC}\pqty{r_{\rm s}} = -\frac{c_0 + c_1r_{\rm s} + c_2r_{\rm s}\ln\pqty{r_{\rm s}} + c_3r_{\rm s}^2}{1 + \pqty{9r_0c_3 / 2\sqrt{3}\pi} r_{\rm s}^4},
\end{equation}
where $\{c_i\}$ are optimized parameters.
These are plotted in Fig.~\ref{fig:softened_dipoles:xc} with the exponent of decay with $r_{\rm s}$ visualized through the logarithmic gradient $D$ in the inset axes.
This verifies that the XC energy tends towards a constant in the $r_{\rm s} \rightarrow 0$ limit, and tends towards the $E\propto 1 / r_{\rm s}^2$ behavior in the low interaction limit to cancel the erroneous decay introduced by the Hartree energy as seen in Eq.~\eqref{eq:hartree_energy}.
Fitted parameters for this model can be found in Table~\ref{tab:softened_xc:fit}.
After fitting, in the paramagnetic case, the $\chi^2$ per degree of freedom is $44.1$ and the RMS error is $5.97\times 10^{-2}$ a.u.*/atom.
The ferromagnetic case exhibited a $\chi^2$ of $540$ per degree of freedom, and an RMS error of $0.131$ a.u.*/atom.

\begin{table}
    \centering
    \caption{Fitting parameters for the XC energy parametrization from Eq.~(\ref{eq:xc_pade-like}) optimized for para- and ferromagnetic softened dipolar systems.  All parameters are in a.u.*}
    \begin{tabular}{ld{8}d{8}}
        \hline\hline
        Param. & \multicolumn{1}{c}{Paramagnetic} & \multicolumn{1}{c}{Ferromagnetic} \\
        \hline
        $c_0$ & 1.89138001  & 1.71788947   \\
        $c_1$ & -4.58498811 & -10.90764358 \\
        $c_2$ & -1.44118629 & -4.56362331  \\
        $a_3$ & 4.74305486  & 16.87599053  \\ \hline\hline
    \end{tabular}%
    \label{tab:softened_xc:fit}
\end{table}

\section{Bare dipolar interaction results}%
\label{sec:bare}

\subsection{Backflow corrections}

In this section results for the bare ($r_0=0$) dipolar interaction are presented, analyzed, and compared to existing data.
Backflow corrections for paramagnetic and ferromagnetic bare dipolar fluids are provided in Table~\ref{tab:backflow_corrections}. They are extrapolated to infinite system size using Eq.~(\ref{eq:energy_extrapolation}).
These corrections are then applied to SJ-DMC results extrapolated to the thermodynamic limit.
The extrapolated SJ-DMC energies per atom without backflow are given alongside their backflow corrected versions in Table~\ref{tab:sj_corrected_energies}.
Due to the divergent nature of the short-range interaction of bare-dipolar fluids and presence of nodes at coalescence points, backflow is a useful technique to employ to reduce nodal surface error.

\begin{table*}[htbp!]
    \centering
    \caption{\label{tab:backflow_corrections} Backflow corrections $B_N$ in bare ($r_0=0$) dipolar fluids calculated for eleven interaction strengths, and their values extrapolated to the thermodynamic limit using Eq.~(\ref{eq:energy_extrapolation}). System sizes were chosen from those proving possible to optimize, and such that the range of $1/N$ was large, reducing the quasirandom finite-size error introduced by this extrapolation.}
    \begin{tabular}{d{3}d{5}e{5}e{5}e{6}e{6}d{6}e{5}}
        \hline\hline
        \multicolumn{1}{l}{\multirow{2}{*}{$d^2$ (a.u.)}} & \multicolumn{3}{c}{Paramagnetic $B_N$ ($10^{-3}$ a.u./atom)} & \multicolumn{4}{c}{Ferromagnetic $B_N$ ($10^{-3}$ a.u./atom)} \\
         & \multicolumn{1}{c}{$N=26$} & \multicolumn{1}{c}{$N=42$} & \multicolumn{1}{c}{$N=\infty$} & \multicolumn{1}{c}{$N=25$} & \multicolumn{1}{c}{$N=37$} & \multicolumn{1}{c}{$N=57$} & \multicolumn{1}{c}{$N=\infty$} \\
        \hline
        0.025 & -2.82(2)  & -2:.50(2)  & -2:.20(5) &         & -0:.006(3) &           &           \\
        0.05  & -3.85(3)  & -3:.47(2)  & -3:.11(5) &         & -0:.002(6) & -0.005(3) & 0:.006(9) \\
        0.1   & -5.44(4)  & -4:.88(3)  & -4:.36(7) &         & -0:.04(1)  & -0.051(7) & -0:.06(2) \\
        0.25  & -8.78(7)  & -7:.24(4)  & -5:.8(1)  &         & -0:.17(2)  & -0.17(1)  & -0:.16(4) \\
        0.5   & -11.39(9) & -10:.37(6) & -9:.4(1)  &         & -0:.45(4)  & -0.42(2)  & -0:.38(7) \\
        1.0   & -15.0(1)  & -13:.87(6) & -12:.8(2) &         & -1:.11(6)  & -1.04(3)  & -0:.96(9) \\
        2.5   & -20.9(2)  & -18:.5(1)  & -16:.2(3) &         & -2:.4(1)   & -2.61(6)  & -2:.8(2)  \\
        5.0   & -25.6(3)  & -22:.3(1)  & -19:.1(4) &         & -4:.0(2)   & -4.5(1)   & -5:.1(3)  \\
        10.0  & -29.2(5)  & -31:.6(2)  & -33:.9(6) &         & -4:.1(4)   & -6.8(2)   & -9:.8(6)  \\
        25.0  & -32.3(5)  & -98:.8(5)  & -162:(1)  & -11:(1) & -19:.2(8)  &           & -29:(2)   \\
        50.0  & -41.1(5)  & -234:(4)   & -418:(7)  &         & -75:(1)    &           &           \\ \hline\hline
    \end{tabular}
\end{table*}

\begin{table*}[htbp!]
    \centering
    \caption{\label{tab:sj_corrected_energies} Backflow-corrected DMC energies per atom of bare ($r_0=0$) dipolar fluids in the thermodynamic limit.}
        \begin{tabular}{d{3}d{8}d{8}d{9}d{8}}
            \hline\hline
            \multicolumn{1}{l}{\multirow{2}{*}{$d^2$ (a.u.)}} & \multicolumn{2}{c}{Paramagnetic} & \multicolumn{2}{c}{Ferromagnetic} \\

            & \multicolumn{1}{c}{$E$ (a.u./atom)} & \multicolumn{1}{c}{$E + B$ (a.u./atom)} & \multicolumn{1}{c}{$E$ (a.u./atom)} & \multicolumn{1}{c}{$E + B$ (a.u./atom)} \\
            \hline
            0.025 & 0.71717(3) & 0.71497(5) & 1.043548(1) &             \\
            0.05  & 0.78910(5) & 0.78599(8) & 1.085102(2) & 1.08511(1)  \\
            0.1   & 0.90430(4) & 0.89995(8) & 1.164622(4) & 1.16456(2)  \\
            0.25  & 1.18029(6) & 1.1745(1)  & 1.386644(8) & 1.38649(4)  \\
            0.5   & 1.5655(2)  & 1.5561(2)  & 1.72758(1)  & 1.72721(7)  \\
            1.0   & 2.24272(2) & 2.2299(2)  & 2.356513(6) & 2.35555(9)  \\
            2.5   & 4.0121(4)  & 3.9959(4)  & 4.07085(5)  & 4.0680(2)   \\
            5.0   & 6.6828(2)  & 6.6637(5)  & 6.70182(8)  & 6.6967(3)   \\
            10.0  & 11.6637(6) & 11.6299(9) & 11.6406(2)  & 11.6308(6)  \\
            25.0  & 25.6950(4) & 25.553(1)  & 25.5963(4)  & 25.567(2)   \\
            50.0  & 48.179(3)  & 47.776(8)  & 47.9123(7)  &               \\ \hline\hline
        \end{tabular}
\end{table*}

\subsection{Relationship between $E$ and $d^2$}

A fitting function for the total energy per atom against interaction strength is
\begin{equation}%
\label{eq:bare:fitting_function}
    E\pqty{d^2} = p_0 + p_1 d + p_2 d^{3/2} + \frac{v_{\text{TR}}d^2}{2} + l_0 \ln\pqty{l_1 d^2 + 1},
\end{equation}
where $\{p_n\}$ and $\{l_n\}$ are free parameters (whose optimized values are reported in Table~\ref{tab:bare:fit}) and $v_{\text{TR}}=1.597017$ a.u.\ is the Madelung constant of a triangular lattice at $d^2=1$ a.u., which provides the strong interaction asymptote.
We note that when treating the constant $v_{\text{TR}}$ as a numerically optimizable parameter it displays strong agreement with the calculated triangular Madelung constant, being 1.596185 a.u.
In fitting this function, it may be tempting to bind the noninteracting limit ($p_0$) to the known value for the infinite system size noninteracting energy per atom, being $0.5$ a.u./atom for paramagnetic systems, and 1 a.u./atom for ferromagnetic.
However, considering that the wave function acquires a node at coalescence points between distinguishable particles for any $d^2>0$, it follows that the total energy per atom is discontinuous at $d^2=0$ for paramagnetic fluids.
In the ferromagnetic case, $p_0$ has been fixed to 1~a.u.\ since no such discontinuity exists.

\begin{figure}[htbp!]
    \centering
    \subfloat[]{%
        \includegraphics[width=0.95\columnwidth]{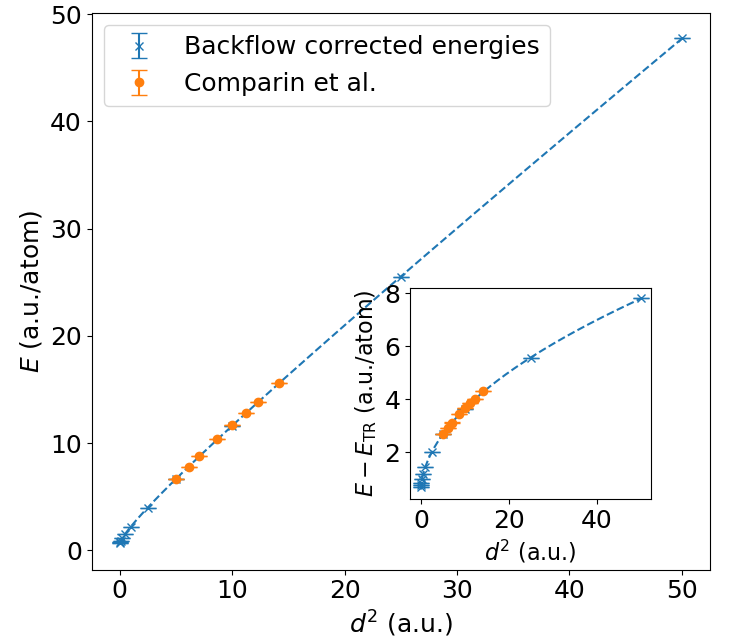}
    } \\
    \subfloat[]{%
        \includegraphics[width=0.95\columnwidth]{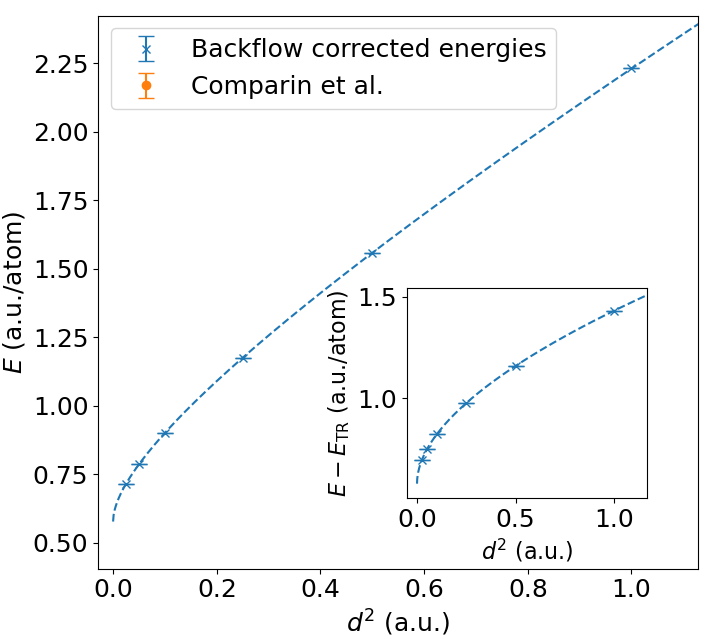}
    }
    \caption{Paramagnetic bare ($r_0 = 0$) DMC energy per atom against interaction strength $d^2$. These results have been compared with Comparin \textit{et al.}~\cite{Comparin2019}. The inset shows the DMC energy per atom relative to the energy per atom of a triangular lattice $E_{\rm TR}$.}%
    \label{fig:bare_E_vs_d2_relationship}
\end{figure}

\begin{figure}[htbp!]
    \centering
    \subfloat[]{%
        \includegraphics[width=0.95\columnwidth]{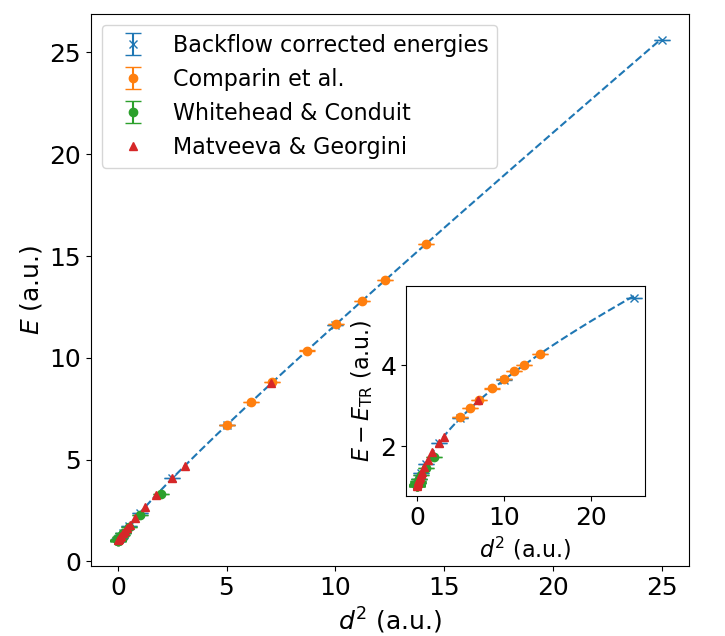}
    } \\
    \subfloat[]{%
        \includegraphics[width=0.95\columnwidth]{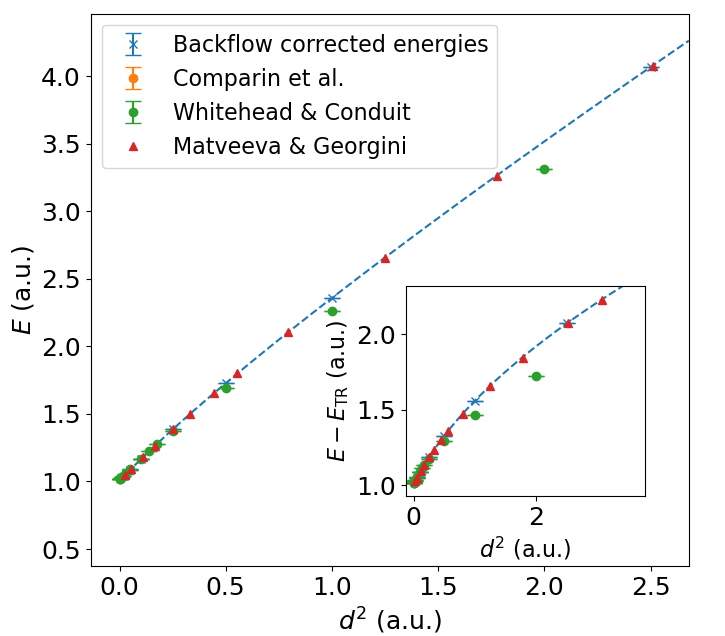}
    }
    \caption{Ferromagnetic bare ($r_0 = 0$) DMC energy per atom against interaction strength $d^2$. Also shown are the results of Comparin \textit{et al.}~\cite{Comparin2019}, Matveeva and Georgini~\cite{Matveeva2012}, and Whitehead and Conduit~\cite{Whitehead2016}. The inset shows the DMC energy per atom relative to the energy per atom of a triangular lattice $E_{\rm TR}$.}%
    \label{fig:bare_E_vs_d2_relationship_ferro}
\end{figure}

The error bars in $E+B$ reported in Table~\ref{tab:sj_corrected_energies} measure QMC error only, and do not account for noise introduced by finite-size extrapolation, especially due to backflow corrections.
This leads to high $\chi^2$ per d.o.f.\ values compared to the fit's RMS error.
The $\chi^2$ per d.o.f.\ is $29.63$ in the ferromagnetic fit and $7.372$ in the paramagnetic fit.
In ferromagnetic systems, the fit produced yields a RMS error of $0.087447$ a.u.\ and in paramagnetic systems this is $0.057599$ a.u.

\begin{table}
    \centering
    \caption{Fitting parameters for the total energy per atom parametrization from Eq.~(\ref{eq:bare:fitting_function}) optimized for para- and ferromagnetic systems. Note that the constant parameter $p_0$ is fixed to 1 a.u.\ in the ferromagnetic case.  All parameters are in a.u.}
    \begin{tabular}{ld{9}d{9}}
        \hline\hline
        Param. & \multicolumn{1}{c}{Paramagnetic} & \multicolumn{1}{c}{Ferromagnetic} \\
        \hline
        $p_0$ & 0.578231131 &             \\
        $p_1$ & 0.692715747 & -0.04325489 \\
        $p_2$ & 0.077518536 & 0.334256760 \\
        $l_0$ & 0.310582528 & 0.350636075 \\
        $l_1$ & 0.302784559 & 1.157897177 \\ \hline\hline
    \end{tabular}%
    \label{tab:bare:fit}
\end{table}

Paramagnetic and ferromagnetic dipolar fermionic systems have been studied previously using QMC simulation by Comparin \textit{et al.}~\cite{Comparin2019}.
However, our QMC methodology differs in a number of regards.
Firstly, as discussed in Appendix~\ref{app:qmc:short_range}, the divergence in local energy in our calculations has been minimized by imposing beyond-leading-order short-range behavior for both paramagnetic and ferromagnetic fluids.
This has led to reduced noise random error in our ferromagnetic fluid energies.
Secondly, finite-size effects in our results are controlled through extrapolation to the thermodynamic limit, whereas in Comparin \textit{et al.}, finite-size errors are reduced by considering a singular large system size ($N=121$ for ferromagnetic and $N=122$ for paramagnetic fluids).

It should be noted that we find no evidence of a Bloch phase transition in backflow corrected energies, with paramagnetic energies being consistently more favorable throughout the considered range of $d^2$ values.
However, we do find one in just the SJ energies, between $d^2=5.0$ and $d^2=10.0$, indicating that the location of such a transition is highly sensitive to nodal surface error in fixed-node DMC.\@
This is consistent with findings noted by Comparin \textit{et al.}, who find no evidence of itinerant ferromagnetism within their range of results.

Data from Comparin \textit{et al.}\ are in good agreement with our results, concurring with our fitted function to at least three significant figures.
In general their energies are slightly lower than ours in the low $d^2$ region, and become slightly higher than ours at high $d^2$, with the crossover taking place at $d^2\approx7$ a.u.
Comparin \textit{et al.}\ examine the QMC energy at $d^2 = 11.2100$ a.u.\ with multiple wave function forms, including iterative backflow and an explicit three-body backflow term. Their lowest energy using any of these methods is $E = 12.775(1)$ a.u./atom in the paramagnetic case and $12.7871(3)$ a.u./atom in the ferromagnetic case.
Our interpolated DMC values for this $d^2$ using the fitting function defined in Eq.~(\ref{eq:bare:fitting_function}) are $E = 12.7776$ and $12.7791$ a.u./atom for the paramagnetic and ferromagnetic cases, respectively.
This is further backed up by the data shown in Fig.~\ref{fig:bare_E_vs_d2_relationship_ferro}, where our ferromagnetic energies are consistently slightly lower than those of Comparin \textit{et al.}, probably due to our imposition of higher-order exact short range behavior on the trial wave function, see Appendix~\ref{app:qmc:short_range}.

For comparison with other studies of ferromagnetic systems of dipolar fermions, results calculated by Whitehead and Conduit~\cite{Whitehead2016}, and Matveeva and Georgini~\cite{Matveeva2012} are displayed in Fig.~\ref{fig:bare_E_vs_d2_relationship_ferro}.
Results from Matveeva and Georgini were obtained through comparison work done by Whitehead and Conduit.
Our results display very good agreement with those of Matveeva and Georgini.
The energy data of Whitehead and Conduit differ significantly from ours, due to the fact that they did not extrapolate their results to the thermodynamic limit, because their work was focused on pseudopotential development.

\section{Conclusions}%
\label{sec:conclusions}

We have calculated DMC energies of homogeneous Fermi fluids of dipolar atoms as a necessary precursor to the simulation of inhomogeneous systems of dressed Rydberg atoms using DFT\@.
Energies for a large variety of interaction strengths with an experimentally derived softening parameter $r_0$ for $^{43}$Ca have been calculated for both para- and ferromagnetic 2D dipolar fermion fluids.
An interpolation formula $E(r_{\rm s})$ was fitted to these data to provide a parametrization that could be used for future DFT calculations.

To our knowledge these are the first QMC calculations aimed to model the homogeneous gas of dressed Rydberg atoms according to their observed behavior. As such they provide a tool to explore homogeneous cold atom traps for this kind of interaction as well as future DFT development. Results for the bare dipolar interaction (limit of no softening parameter) were compared with existing literature, finding good agreement.
Extrapolations of backflow corrections to the thermodynamic limit constituted estimates for the fixed-node error present in our SJ-DMC calculations.
An interpolation formula $E\pqty{d^2}$ was fitted to our data.
Backflow corrections were found to be much larger at high interaction strengths and in paramagnetic systems.
Short range divergences in local energy were mitigated through the Jastrow factor, resulting in less noise and more precision in ferromagnetic calculations than previous studies.
We found no region of stability for itinerant ferromagnetism within the parameter spaces considered for both softened and bare dipolar calculations.
Future work would include simulation of a four state system relevant to spin 3/2-atoms as well as QMC calculations on Wigner crystals aiming to determine the $r_{\rm s}$ value for crystallization, as has been done in the bare dipolar case by Matveeva and Georgini~\cite{Matveeva2012}.
It could facilitate DFT studies of cold atom systems of dressed Rydberg atoms, permitting greater computational insight into quantum simulators.

In summary, a comprehensive dataset for homogeneous 2D dipolar systems, particularly those from dressed Rydberg atoms, has been created, enabling the use of DFT calculations on inhomogeneous systems.

\begin{acknowledgments}
    Our calculations were performed using Lancaster University's High End Computing (HEC) cluster.
    We would also like to thank Pablo L\'{o}pez R\'{\i}os, Leon Petit, and Gareth J.\ Conduit for valuable conversations regarding this work.
\end{acknowledgments}

\appendix

\section{Short-range behavior}%
\label{app:qmc:short_range}

It is critical to control divergences in local energies arising from the coalescence of dipoles.
This Appendix discusses the approach taken in QMC to mitigate these divergences.
The leading-order short-range behavior of the Jastrow factor for the bare dipolar fluid may be determined by requiring that the local kinetic energy provides an opposing divergence to cancel the $1/r^3$ divergence in energy. Beyond-leading order short-range behavior can be imposed by eliminating as many divergent terms from the local energy as possible.
Let $\hat{H}'$ be the component of the Hamiltonian due to the relative motion of a coalescing pair of dipolar atoms, with $\mu$ being the reduced mass of the coalescing pair.  Taking the form of the trial wave function used in Eq.~(\ref{eq:trial_wavefunction}), the contribution of two such atoms to the local energy is
\begin{equation}
    \begin{split}
        E_L^{\prime} =& \frac{\hat{H}^{\prime}\psi_{\rm T}}{\psi_{\rm T}} =
        -\frac{1}{2\mu}\Biggl(
            \frac{\nabla^2\exp\pqty{J}}{\exp\pqty{J}} \\
                      &+ 2\frac{\nabla\exp\pqty{J}}{\exp\pqty{J}}\cdot\frac{\nabla S}{S}
        + \frac{\nabla^2 S}{S}
        \Biggr) + \frac{d^2}{r^3}.
    \end{split}
\end{equation}
For distinguishable atoms here the $\pqty{\nabla S}/ S$ factor depends upon the entire configuration $\vb{R}$ and not just the coalescing atoms.
Therefore the $\nabla S \cdot \nabla \exp\pqty{J} / \psi_{\rm T}$ term cannot be canceled.
We subsequently require
\begin{equation}
    -\frac{1}{2\mu\exp\pqty{J}}\pqty{\dv[2]{\exp\pqty{J}}{r}+\frac{1}{r}\dv{\exp\pqty{J}}{r}}+\frac{d^2}{r^3}=0,
\end{equation}
which via substitution can be re-expressed as a zeroth order Bessel equation, yielding the approximate solution
\begin{equation}
    \exp\pqty{J} = K_0\pqty{a_{d2} / \sqrt{r}} \approx \exp\pqty{-a_{d2} / \sqrt{r}},
\end{equation}
where $a_{d2} = \sqrt{8d^2\mu}$.
This removes the leading order $O\pqty{r^{-3}}$ divergence in the local energy, but as stated earlier cannot remove the leading order behavior in $\nabla S\cdot\nabla\exp\pqty{J}/\psi_{\rm T}$ which has a divergence of $O\pqty{r^{-3/2}}$.
Additionally, this divergence disappears upon spherically averaging the contribution to the local energy such that the remaining divergence has only a small prefactor in most configurations.
For indistinguishable atoms, on the other hand, the following condition is required
\begin{equation}
    -\frac{1}{2\mu\exp\pqty{J}}\pqty{\dv[2]{\exp\pqty{J}}{r}+\frac{3}{r}\dv{\exp\pqty{J}}{r}}+\frac{d^2}{r^3}=0.
\end{equation}
The solution to which via substitution can be re-expressed as a second order Bessel equation with the approximate solution
\begin{equation} \exp\pqty{J} = r^{-1} K_2\pqty{a_{d2} / \sqrt{r}} \approx \exp\pqty{-a_{d2} / \sqrt{r}}.
\end{equation}
This leaves a $O\pqty{r^{-1/2}}$ divergence between all atoms in a ferromagnetic system.

The form of the two-body Jastrow term that imposes leading order behavior is $u_{\rm A}\pqty{r} = \ln\bqty{K_0\pqty{a_{d2}/\sqrt{r}}}$ for distinguishable pairs and $u_{\rm A}\pqty{r} = \ln\bqty{K_2\pqty{a_{d2}/\sqrt{r}}}$ for indistinguishable pairs.
In order to maintain computational viability, Jastrow terms are truncated smoothly to a radius of $L_u$ to account for the more relevant short-range correlation.
This is achieved through transforming $u_{\rm A}\pqty{r}$ to
\begin{equation}
    u\pqty{r} = \bqty{u_{\rm A}\pqty{r} - u_{\rm A}\pqty{L_u}}u_{\rm C},
\end{equation}
where
\begin{equation}
    u_{\rm C} = \pqty{1 - \frac{r^3}{L_u^3}}^{C - 1}\Theta\pqty{L_u - r}.
\end{equation}
Here $C$ is an integer cutoff exponent and $\Theta$ is a Heaviside step function.
The cutoff function $u_{\rm C}$ deviates from 1 by $O\pqty{r^3}$ at short range, which ensures it does not combine with the leading order $r^{-1/2}$ behavior to introduce new divergent terms into the local kinetic energy.

\section{Hartree{--}Fock theory}%
\label{app:qmc:hartree-fock}

\subsection{Kinetic energy}%
\label{app:qmc:hartree-fock:kinetic}

The kinetic energy per atom of an infinite, homogeneous fluid of
uncorrelated fermionic atoms is $E_{\rm K} = 1/r_{\rm s}^2$ for
ferromagnetic systems and $E_{\rm K} = 1/(2r_{\rm s}^2)$ for
paramagnetic systems.

\subsection{Hartree energy}%
\label{app:qmc:hartree_energy}

The Hartree energy per atom of a homogeneous softened dipolar fluid is
\begin{eqnarray} E_{\rm H} & = & \frac{1}{2N}\int_A \int_A n^2 \sum_{{\bf R}_{\rm s}} v_{\rm dd}(|{\bf r}_1-{\bf r}_2 +{\bf R}_{\rm s}|) \, {\rm d}^2{\bf r}_1 \, {\rm d}^2{\bf r}_2 \nonumber \\ & = & \frac{n}{2} \int_0^\infty 2\pi r \, v_{\rm dd}(r) \, {\rm d}r \nonumber \\ & = & \frac{2\sqrt{3}\pi d^2}{9r_0 r_{\rm s}^2}. \label{eq:hartree_energy}\end{eqnarray}

Note that the Hartree energy diverges as $r_0 \to 0$; hence we cannot define an XC energy in this limit.  Furthermore, even for finite $r_0$, the Hartree energy provides an inadequate description of the interaction energy at low density ($r_{\rm s} \gg r_0$). At low density correlation prevents the particles approaching each other, so that the interaction energy falls off as $1/r_{\rm s}^3$ rather than the $1/r_{\rm s}^2$ behavior predicted by Hartree theory.

In practical cold atom systems formed with optical tweezers,  it is typically the case that $r_{\rm s}\lesssim r_{0}$, so this is not a major issue for such systems.
As determined in Ref.~\onlinecite{Hague2017}, for $^{43}$Ca, $r_{\rm s}/r_0$ values are 0.6938 where $n_{\text{Ryd}}=32$ and 0.5884 where $n_{\text{Ryd}}=38$.
For $^{87}$Sr they are 0.7263 where $n_{\text{Ryd}}=32$ and 0.5896 where $n_{\text{Ryd}}=38$.
For $^{25}$Mg they are 0.7005 where $n_{\text{Ryd}}=32$ and 0.5939 where $n_{\text{Ryd}}=38$.

\subsection{Exchange energy}%
\label{app:qmc:exchange_energy}

The exchange energy per atom due to spin-up atoms is
\begin{equation} 
    E_{{\rm X}\uparrow} = -\frac{1}{2{(2\pi)}^4 n} \iint_{{\bf k}:k<k_{{\rm F}\uparrow} \atop {\bf k}':k'<k_{{\rm F}\uparrow}} v_{\rm d}(|{\bf k}-{\bf k}'|) \, {\rm d}^2{\bf k} \, {\rm d}^2{\bf k}',
\end{equation}
where
\begin{eqnarray}
    v_{\rm d}(q) & = & \int v_{\rm dd}(r) \exp(-i{\bf q}\cdot{\bf r}) \, {\rm d}^2{\bf r} \nonumber \\
    & = & 2\pi \int_0^\infty v_{\rm dd}(r) J_0(qr) r \, {\rm d}r
\end{eqnarray}
is the Fourier transform of Eq.\ (\ref{eq:dipole_dipole_interaction}) and $k_{{\rm F}\uparrow}=\sqrt{4\pi n_\uparrow}$ is the Fermi wavevector for spin-up atoms.
$J_0$ is the zeroth-order Bessel function.
Although $v_{\rm d}(q)$ cannot be found analytically, it can be computed numerically on a logarithmic grid by a fast Hankel transform, and then evaluated at any $q$ by cubic spline interpolation.
The exchange energy per atom can be reduced to a one-dimensional integral that can be performed numerically:
\begin{equation}
    E_{{\rm X}\uparrow} = - {\left(\frac{r_{\rm s} k_{{\rm F}\uparrow}^2}{4 \pi}\right)}^2 \int_0^2 a_1(\tilde{q}) v_{\rm d}(\tilde{q}k_{{\rm F}\uparrow}) \tilde{q} \, {\rm d}\tilde{q},
\end{equation}
where
\begin{equation}
    a_1(\tilde{q}) = 2 \cos^{-1}\left(\frac{\tilde{q}}{2}\right) - \tilde{q} \sqrt{1-{\left(\frac{\tilde{q}}{2}\right)}^2}
\end{equation}
is the area of overlap between two circles of unit radius whose centers are a distance $\tilde{q}$ apart.
A similar expression is obtained for the exchange energy per atom due to spin-down atoms.
The total exchange energy per atom is $E_{\rm X}=E_{{\rm X}\uparrow}+E_{{\rm X}\downarrow}$.

In the low-density limit ($r_{\rm s}\to \infty$), $v_{\rm d}(\tilde{q}k_{{\rm F}\uparrow})$ can be approximated by $v_{\rm d}(q=0)$, in which case the exchange energy per atom reduces to $E_{\rm X} = -E_{\rm H}/2$ for a paramagnetic fluid and $E_{\rm X} = -E_{\rm H}$ for a ferromagnetic fluid.

In the high density limit ($r_{\rm s}\to 0$), $\int_{{\bf k}':k'<k_{{\rm F}\uparrow}} e^{-i{\bf k}'\cdot {\bf r}} \, d{\bf k}' \to {(2\pi)}^2\delta({\bf r})$ and hence the exchange energy per atom tends to a constant, $E_{\rm X} = -v_{\rm dd}(0)/2$, independent of spin polarization.

The lack of a straightforward expression for the exchange energy at finite $r_{\rm s}$, together with the fact that the erroneous $O(r_{\rm s}^{-2})$ leading-order contribution to the interaction energy in Hartree{--}Fock theory at large $r_{\rm s}$ is only eliminated for the fully ferromagnetic fluid, suggests there is no particular advantage to parameterizing the correlation energy of the softened interaction as opposed to the XC energy.

\section{Finite-size effects in dipolar atomic gases}%
\label{app:qmc:finite_size}

When the effects of periodic images are included, the interaction between two dipolar atoms with separation ${\bf r}$ within the simulation cell is
\begin{equation}
    v_{\rm d}({\bf r}) \equiv \sum_{{\bf R}_{\rm s}}
    v_{\rm dd}(|{\bf r}+{\bf R}_{\rm s}|),
\end{equation}
where $v_{\rm dd}(r)$ is the softened pairwise dipolar interaction in Eq.\ (\ref{eq:dipole_dipole_interaction}).
The Madelung constant is the sum of interactions between one dipolar atom and all its periodic images:
\begin{equation}
    v_{\rm M} \equiv \sum_{{\bf R}_{\rm s} \neq {\bf 0}} v_{\rm dd}(R_{\rm s}) = v_{\rm d}({\bf r}={\bf 0}) - v_{\rm dd}(0).
\end{equation}

Let
\begin{eqnarray}
    v_{\rm d}({\bf G}) & \equiv & \int_A v_{\rm d}({\bf r}) \exp(-i{\bf G} \cdot {\bf r}) \, {\rm d}^2{\bf r} \\
    & = & \int_\text{All space} v_{\rm dd}(r) \exp(-i{\bf G} \cdot {\bf r}) \, {\rm d}^2{\bf r}
\end{eqnarray}
be the Fourier transform of $v_{\rm d}({\bf r})$, where $\{{\bf G}\}$ are the simulation-cell reciprocal lattice points.
$v_{\rm d}({\bf G})$ is clearly independent of system size. 
Note that $v_{\rm d}({\bf r}={\bf 0})=(1/A) \sum_{\bf G} v_{\rm d}({\bf G})$.

In reciprocal space, the expectation value of the potential energy per particle and the long-range two-body Jastrow contribution to the kinetic energy per particle are, respectively,
\begin{eqnarray}
    \langle \hat{V} \rangle & = & \frac{1}{2} \left( v_{\rm M} + \frac{1}{A} \sum_{\bf G} v_{\rm dd}({\bf G}) \left[ S({\bf G})-1\right] \right) + E_{\rm H} \nonumber \\
    & = & \frac{1}{2} \left( -v_{\rm dd}(0) +\frac{1}{A} \sum_{\bf G} v_{\rm d}({\bf G}) S({\bf G}) \right) + E_{\rm H} \label{eq:V} \\
    \langle \hat{T}_{\rm lr} \rangle & = & \frac{1}{4A} \sum_{\bf G} G^2 u({\bf G}) S^*({\bf G}) - \frac{1}{4A} \sum_{\bf G} G^2 u({\bf G}), \label{eq:T_lr}
\end{eqnarray}
where $u({\bf G})$ is the Fourier transform of the two-body Jastrow factor and
\begin{equation}
    S({\bf G})=\left[ \langle \hat{\rho}({\bf G}) \hat{\rho}^*({\bf G})\rangle - \rho({\bf G}) \rho^*({\bf G})\right]/N
\end{equation}
is the static structure factor, where $\hat{\rho}({\bf G})=\sum_i \exp(-i{\bf G}\cdot \hat{\bf r}_i)$ is a Fourier component of the density operator and $\rho({\bf G})=\langle \hat{\rho}({\bf G})\rangle$~\cite{Chiesa2006,Drummond2008,Holzmann2016}.
Note that in real space,
\begin{eqnarray}
    S({\bf r}) & = & \delta({\bf r}) + \frac{1}{N} \int_A \left[\rho_2({\bf r}',{\bf r}+{\bf r}')-\rho({\bf r}')\rho({\bf r}+{\bf r}')\right] \, {\rm d}^2{\bf r}' \nonumber \\
    & \equiv & \delta({\bf r}) + \rho_{\rm xc}({\bf r}),
\end{eqnarray}
where the atomic density is $\rho({\bf r})=\langle \hat{\rho}({\bf r}) \rangle$, $\hat{\rho}({\bf r})=\sum_i \delta({\bf r}-\hat{\bf r}_i)$, and the pair density is $\rho_2({\bf r},{\bf r}')=\langle \sum_{i \neq j} \delta({\bf r}-\hat{\bf r}_i) \delta({\bf r}'-\hat{\bf r}_j) \rangle$.
Note that $\int \rho_{\rm xc}({\bf r}) \, {\rm d}^2{\bf r}=-1$.
Hence $S({\bf r})$ can be interpreted as the atomic density due to an atom at the origin and the surrounding depletion in atomic density around that atom known as the exchange-correlation hole $\rho_{\rm xc}({\bf r})$.

Assuming the structure factor is well-converged with respect to system size, the primary source of finite-size error in the energy per particle is the difference between summation over discrete reciprocal lattice points ${\bf G}$ and integration over ${\bf k}$ in reciprocal space in Eqs.\ (\ref{eq:V}) and (\ref{eq:T_lr}).
By the Poisson summation formula $A^{-1} \sum_{\bf G} f({\bf G}) - {(2\pi)}^{-2}\int f({\bf k}) \, {\rm d}^2{\bf k} = \sum_{{\bf R}_{\rm s}\neq {\bf 0}}f({\bf R}_{\rm s})$ for any function $f({\bf r})$ whose Fourier transform $f({\bf k})$ exists.
In Eq.\ (\ref{eq:V}), this gives a finite-size correction to the potential energy per atom
\begin{eqnarray}
    \Delta V & = & \frac{1}{2} \left[ \frac{1}{{(2\pi)}^2} \int v_{\rm d}({\bf k}) S({\bf k}) \, {\rm d}^2{\bf k} - \frac{1}{A} \sum_{\bf G} v_{\rm d}({\bf G}) S({\bf G}) \right] \nonumber \\
    & = & -\sum_{{\bf R}_{\rm s} \neq {\bf 0}} \int_\text{All space} S({\bf r}) v_{\rm dd}(|{\bf R}_{\rm s}-{\bf r}|) \, {\rm d}^2{\bf r} \nonumber \\
    & \sim & -\sum_{{\bf R}_{\rm s} \neq {\bf 0}} R_{\rm s}^{-5} \sim -N^{-5/2}.
\end{eqnarray}
In the second step we have noted that the inverse Fourier transform of product $v_{\rm d}({\bf G})S({\bf G})$ is the convolution of $v_{\rm dd}({\bf r})$ and $S({\bf r})$, giving an expression for the dipolar potential at each ${\bf R}_{\rm s} \neq {\bf 0}$ due to a dipolar density $S({\bf r})$ at the origin.
In the third line we have noted that $S({\bf r})=\delta({\bf r})+\rho_{\rm xc}({\bf r})$ integrates to zero and that the dipole moment of atomic density $S({\bf r})$ is zero by symmetry; however, in general $S({\bf r})$ will have a quadrupole moment, so that the long-range dipolar potential due to atomic density $S({\bf r})$ goes as $r^{-5}$.

The long-range pairwise dipolar interaction goes as $v_{\rm dd}(r) \sim r^{-3}$ in real space and therefore $v_{\rm d}({\bf k}) \sim k$ at small $k$ in reciprocal space.
Hence the static structure factor goes as $S({\bf k}) \sim k$ and so the two-body Jastrow factor goes as $u({\bf k}) \sim -k^{-1}$ at small $k$ in reciprocal space, i.e., as $-r^{-1}$ at long range in real space~\cite{Gaskell1961,Holzmann2016}.
We have verified that this long-range behavior holds in practice for our optimized trial wave functions, for both the softened and bare dipolar potentials.

The leading-order correction to the finite-size error in the long-range kinetic energy per atom is due to the second term in Eq.\ (\ref{eq:T_lr}), and goes as
\begin{eqnarray}
    \Delta T_{\rm lr} & = & -\frac{1}{4} \left[ \frac{1}{{(2\pi)}^2} \int k^2 u({\bf k}) \, {\rm d}^2{\bf k} - \frac{1}{A} \sum_{\bf G} G^2 u({\bf G}) \right] \nonumber \\
    & = & -\frac{1}{4} \sum_{{\bf R}_{\rm s} \neq {\bf 0}} \nabla^2 u({\bf R}_{\rm s}) \nonumber \\
    & \sim & \sum_{{\bf R}_{\rm s} \neq {\bf 0}} R_{\rm s}^{-3} \sim +N^{-3/2},
\end{eqnarray}
where we have noted that $\nabla^2 r^{-1}=-r^{-3}$ in 2D\@.

Hence the leading-order systematic finite-size correction to the total energy per particle of a 2D homogeneous dipolar gas is due to the long-range kinetic energy, is positive, and goes as $N^{-3/2}$.
For the bare dipolar interaction, the Fourier components of $v_{\rm dd}(r)$ diverge as $r_0 \to 0$; however, the leading-order finite-size error arises from a kinetic-energy contribution that does not explicitly depend on $v_{\rm d}({\bf k})$, and so this finite-size scaling applies to both the softened and bare dipolar interactions.

\section{Error resulting from integral approximation to real-space sum}%
\label{app:qmc:err_sum}

\begin{figure}
    \centering
    \includegraphics[width=0.95\columnwidth]{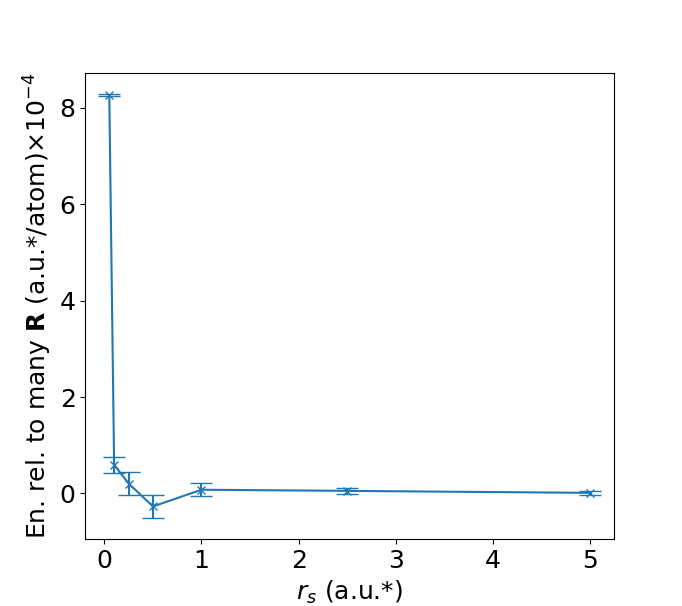}
    \caption{SJ-VMC energies for a $N=97$ ferromagnetic system relative to those calculated with a large number of stars of $\vb{R}_{\rm s}$ are plotted against $r_{\rm s}$}%
    \label{fig:err_sum}
\end{figure}

Slater-Jastrow VMC calculations were performed with $R_c$ set to truncate the sum after the first 119 stars as has been used throughout the rest of the paper.
These were compared to results which truncated the sum after including 1190 stars of $\vb{R}_{\rm s}$.
The resultant effect is sensitive to $N$, which means it can be considered a finite-size effect.
The relative energies are plotted in Fig.~\ref{fig:err_sum}

These results show the error to be negligible for less dense systems which occupy more real space overall.
The data point located at $r_{\rm s} = 0.05$ displays a considerable error when compared with QMC error.
This error is still considerably smaller than the smallest energies considered in this paper's results (the softened XC energies).

\bibliography{dipolar-fermions.bib}

\end{document}